# Reduction of multiple filamentation due to atmospheric turbulence during the propagation of a laser pulse

Alain Bourdier[1], Hassen Ghalila[2,3], Olivier Delage[4]

[1]Department of Physics and Astronomy, The University of New Mexico, Albuquerque, NM, USA.
[2]De Vinci Higher Education, De Vinci Research Center, Paris, France
[3]Département de Physique, Faculté des Sciences de Tunis, Université Tunis El Manar, 2092 El Manar, Tunis, Tunisie
[4]Laboratoire d'Aérologie UMR 5560 université de Toulouse, France

**Abstract**

Multiple filamentation poses a significant challenge for laser pulse propagation in the atmosphere. This article investigates how atmospheric turbulence influences the development of modulational instability, which leads to multiple filamentations. Through various analytical approaches, we demonstrate that the growth rate of this instability decreases when the refractive index exhibits stochastic behavior.

## 1) Introduction

Understanding how laser pulses propagate in the atmosphere is essential for applications such as remote sensing of chemical and biological agents, as well as directed energy systems [1-9]. For instance, the LIDAR (light detection and ranging) technique is an effective tool for investigating atmospheric pollutants.

When the laser power exceeds the critical power $P \gtrsim P_{cr}$, unavoidable beam distortions and refractive index irregularities cause the input beam to break into $N$ filaments, with $N \sim P/P_{cr}$, each carrying approximately the critical power [10,11]. This results in the formation of chaotic filament bundles and an erratic backscattering signal. The impact of multiple filamentation can be either beneficial or detrimental, depending on specific measurements or applications. For example, lightning discharge control requires long, homogeneous plasma channels generated by filaments [12-15]. The backscattered nitrogen fluorescence detected within the filaments exhibits irregular shot to shot variations that cannot be attributed solely to fluctuations resulting in the initial laser pulse, assuming stable laser energy [16]. The beam profile always exhibits perturbations in both intensity and phase within the light field which vary from shot to shot. Multiple filamentation arises as a consequence of this transverse modulation [16-18]. At the initial stage of the beam, certain maxima in energy density are present (Fig.1). These irregularities can be amplified by the Bespalov and Talanov modulational instability [19]. We demonstrate that turbulence inhibits the development of this modulational instability [20]. Due to Kerr effect, that is the nonlinear component $\bar{n}_2 I$ of the refractive index [1], the light rays are confined along regions of positive refractive index gradients, leading to filament formation [8, 21, 22]. Consequently, turbulence can reduce the extent of multiple filamentation.

For many applications, controlling filamentation is a critical issue and managing the modulational instability that leads to multiple filamentation is a challenge. To further demonstrate that the growth rate of the Bespalov and Talanov instability [19] can be reduced by turbulence [20] various methods are employed. Specifically, a stochastic refractive index is introduced into the nonlinear Schrödinger equation to account for turbulence. To show that the instability is well reduced, we first provide a physical explanation for the instability in the form of an enlightening reminder in Appendix A. A mismatch explains the decrease in growth rate of modulational instability when the atmosphere is turbulent. Next, we conduct a simple investigation of the problem. Then, we use the Fokker-Planck equation [23, 24] for a more in-depth analysis. This equation describes the evolution of a particle distribution function under the influence of random forces. We present different approaches, acknowledging that the physical assumptions may vary across them. However, we demonstrate that, regardless of the approach, turbulence systematically minimizes the growth rate of Bespalov and Talanov instability.

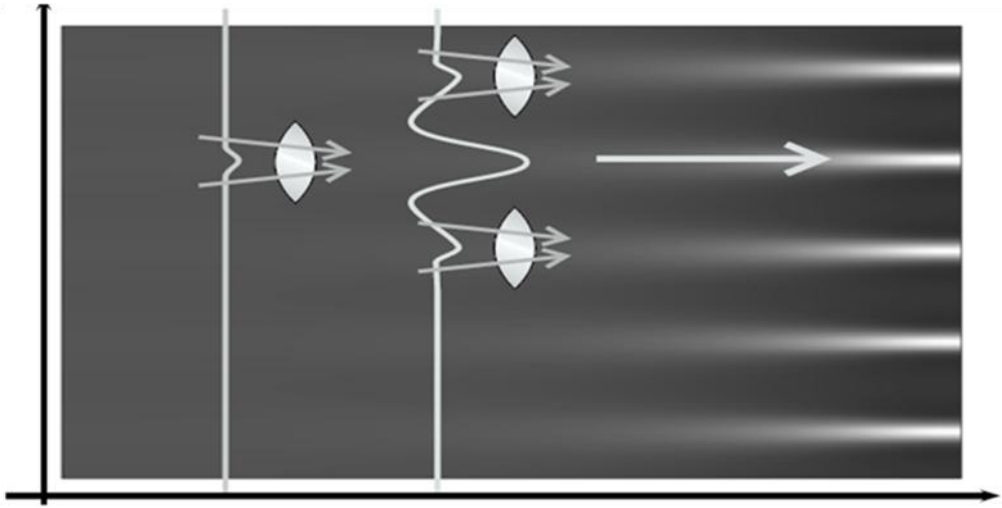

Fig. 1 Multiple filamentation is initiated by the modulational instability.

In short, the purpose of this article is to emphasize that multiple filamentation can be minimized when the atmosphere is turbulent through several analytical approaches. This should result in a beam of light that is almost free from intensity discontinuities.

## 2) Model equations

We are interested in the propagation of an electromagnetic field through a turbulent atmosphere. Its electric field is given by the following expression

$$E\left[\mathbf{r}=(x,y,z),t\right]=\psi(\mathbf{r},t)\,exp\,i(k_0 z - \omega_0 t)\frac{\hat{\mathbf{e}}_x}{2}+cc, \qquad (1)$$

where $\psi(\mathbf{r},t)$ varies slowly in time and $z$ and with $k_0 = n_0\left(\omega_0/c\right)$, where $n_0$ is the space average of the linear refractive index and $\mathbf{r}=(x,y,z)$ the spatial dependence. The total refractive index is $n = n_0 + 2n_2 \langle E^2(\mathbf{r},t)\rangle + \delta n(\mathbf{r})$ where the angular brackets represent a time average and $\delta n(\mathbf{r})$ is a random function characterizing the atmosphere turbulence, the nonlinear index which quantifies Kerr effect is considered here, $n_2$ is the second-order index of refraction [1, 25-27]. The refractive index is assumed to depend only on space as a ray goes through the atmosphere so quickly that it does not change significantly during the crossing.

Applying the Fourier transform to the wave equation, $\partial^2 E/\partial z^2 + \Delta_\perp E - (1/c^2)\partial^2 D/\partial t^2 = 0$ with $D(r,t) = \int_{-\infty}^{t} \varepsilon(t') E(r, t-t') dt'$, gives

$$\frac{\partial^2}{\partial z^2} E(\mathbf{r}, \omega) + \Delta_\perp E(\mathbf{r}, \omega) + \frac{\varepsilon(\omega)}{\varepsilon_0} \frac{\omega^2}{c^2} E(\mathbf{r}, \omega) = 0. \tag{2}$$

The quantity $\varepsilon$ is the atmosphere dielectric constant, $\varepsilon_0$ the permittivity of free space. As the amplitude of the wave is assumed to vary slowly in time, $\psi$ contains no high-frequency components and we consider that $E(\mathbf{r},\omega) \approx (1/2)\psi(\mathbf{r}, \omega - \omega_0) e^{ik_0 z}$. Moreover, the slowly varying amplitude in space approximation is made and $\partial^2 \psi/\partial z^2$ can be dropped [25]. The wave equation, in terms of $\psi$, reads

$$2ik_0 \frac{\partial}{\partial z}\psi + \Delta_\perp \psi + (k^2 - k_0^2)\psi = 0, \tag{3}$$

where $\psi = \psi(\mathbf{r}, \omega - \omega_0)$ and $k(\omega) = \sqrt{\varepsilon(\omega)/\varepsilon_0}\,\omega/c$. As $\langle E^2(\mathbf{r},t)\rangle = (1/2)\psi\psi^* = (1/2)|\psi|^2$, we have now $n = n_0 + n_2|\psi|^2 + \delta n(\mathbf{r})$. The propagation constant k depends on the frequency and intensity of the wave. Here this dependence is described in terms of a series expansion to second order in $\omega - \omega_0$. We have

$$\begin{aligned}k &= k_0 + \Delta k_{NL} + \left(\frac{dk}{d\omega}\right)_{\omega=\omega_0}(\omega-\omega_0) + \frac{1}{2}\left(\frac{d^2k}{d\omega^2}\right)_{\omega=\omega_0}(\omega-\omega_0)^2 + \delta n \frac{\omega_0}{c}, \\ &= k_0 + \Delta k_{NL} + \frac{1}{v_g(\omega=\omega_0)}(\omega-\omega_0) - \frac{1}{2}\left(\frac{1}{v_g^2}\frac{dv_g}{d\omega}\right)_{\omega=\omega_0}(\omega-\omega_0)^2 + \delta n \frac{\omega_0}{c},\end{aligned} \tag{4}$$

in this equation, $v_g$ is the group velocity and $\Delta k_{NL} = (\omega_0/c) n_2 |\psi(\mathbf{r},\omega)|^2$ is the nonlinear contribution to the propagation constant. Then, the equation of propagation is reformulated in the time domain by using the inverse Fourier transform as follows: $\frac{1}{2\pi}\int_{-\infty}^{+\infty} \psi(\mathbf{r}, \omega-\omega_0) e^{-i(\omega-\omega_0)t} d(\omega-\omega_0) = \psi(\mathbf{r},t)$ and thus $\frac{1}{2\pi}\int_{-\infty}^{+\infty} (\omega-\omega_0)^n \psi(\mathbf{r}, \omega-\omega_0) e^{-i(\omega-\omega_0)t} d(\omega-\omega_0) = i^n \frac{\partial^n}{\partial t^n}\psi(\mathbf{r},t)$. Next, a retarded time $\tau$ is introduced: $\tau = t - z/v_g$. The wave equation becomes [25]

$$k_0 \frac{1}{v_g^2} \frac{dv_g}{d\omega} \frac{\partial^2}{\partial \tau^2} \psi_z(x,y,z,t) + 2ik_0 \psi_z + \Delta_\perp \psi + 2k_0^2 \frac{n_2}{n_0} |\psi|^2 \psi = -2 \frac{k_0^2}{n_0} \delta n \psi(\mathbf{r}). \quad (5)$$

Next, it is assumed that the group velocity dispersion can be neglected. Consequently, the following equation describing the pulse propagation in a turbulent atmosphere is derived [20, 25, 28-30]

$$2ik_0 \psi_z [\mathbf{r}=(x,y,z)] + \Delta_\perp \psi(\mathbf{r}) + 2k_0^2 \frac{n_2}{n_0} |\psi|^2 \psi(\mathbf{r}) = -2 \frac{k_0^2}{n_0} \delta n \psi(\mathbf{r}). \quad (6)$$

To predict the amplification of irregularities initially present on the laser wavefront, a radial perturbation of $\psi$ is introduced [19, 20, 25]

$$\psi(\mathbf{r}) = \psi_0(z) + a_1(z) e^{i\mathbf{k}_\perp \cdot \mathbf{r}} + a_{-1}(z) e^{-i\mathbf{k}_\perp \cdot \mathbf{r}}, \quad (7)$$

where $a_1(z)$ and $a_{-1}(z)$ are first-order quantities and $\pm k_\perp$ are the transverse components of the wavevector of the off-axis modes. We begin by assuming that fluctuations in the refractive index are in the form

$$\delta n(z) = \delta n_0 \cos \varphi(z), \quad (8)$$

where $\delta n_0$ is a first-order constant and $\varphi$ a stochastic angle. Throughout this article it is considered that $\langle \cos \varphi(z) \rangle = 0$ where angular brackets represent an average.

To find a solution to the wave equation [Eq. (6)], we first assume that $\psi_0$ satisfies

$$\frac{d\psi_0(z)}{dz} - i \frac{\omega_0}{c} \psi_0 \left( n_2 |\psi_0|^2 + \delta n_0 \cos \varphi \right) = 0. \quad (9)$$

The solution is $\psi_0(z) = \bar{\psi} \exp\left[ i \left( \gamma z + \frac{\omega_0}{c} \delta n_0 \int_0^z \cos \varphi \, dz \right) \right]$ where $\gamma = \frac{\omega_0}{c} n_2 \bar{\psi}^2$ and $\bar{\psi}$ is a real constant which is a zero-order quantity. We consider $\xi = \delta n_0 / (n_2 \bar{\psi}^2)$ is small and neglect the stochastic integral.

Always with the aim of satisfying the wave equation, the expansion for $\psi(\mathbf{r})$ is considered. The propagation equation is satisfied by setting to zero the terms in $e^{i\mathbf{k}_\perp \cdot \mathbf{r}}$ and in $e^{-i\mathbf{k}_\perp \cdot \mathbf{r}}$ [20, 25]. Thus, we obtain the wave equation for the off-axis modes

$$2ik_0 \frac{\partial a_1}{\partial z} - k_\perp^2 a_1 = -k_0^2 \frac{2n_2}{n_0} \bar{\psi}^2 \left( 2a_1 + a_{-1}^* e^{2i\gamma z} \right) - \frac{2k_0^2}{n_0} \delta n_0 \cos \varphi \, a_1,$$

$$2ik_0 \frac{\partial a_{-1}}{\partial z} - k_\perp^2 a_{-1} = -k_0^2 \frac{2n_2}{n_0} \bar{\psi}^2 \left( 2a_{-1} + a_1^* e^{2i\gamma z} \right) - \frac{2k_0^2}{n_0} \delta n_0 \cos \varphi \, a_{-1}. \quad (10)$$

Letting $a_{\pm 1} = \mathscr{A}_{\pm 1} e^{i\gamma z}$, $\beta = k_\perp^2 / 2k_0$ and $\chi = k_0 \delta n_0 / n_0$, this set of equations can be written using the form below

$$\frac{d}{dz}\begin{pmatrix} \mathscr{A}_1 \\ \mathscr{A}_{-1}^* \end{pmatrix} = M_\varphi \begin{pmatrix} \mathscr{A}_1 \\ \mathscr{A}_{-1}^* \end{pmatrix}, \tag{11}$$

with

$$M_\varphi = \begin{bmatrix} i(\gamma - \beta + \chi \cos\varphi) & i\gamma \\ -i\gamma & -i(\gamma - \beta + \chi \cos\varphi) \end{bmatrix}. \tag{12}$$

We have previously applied this equation to predict the impact of atmospheric turbulence on the modulational instability [20].

### 3) Calculation of the growth rate of modulational instability in the absence turbulence. Introduction of a mismatch that accounts for the decrease in the growth rate of modulational instability when the atmosphere is turbulent.

We seek the eigen values of matrix (12) in which $\chi = 0$. We solve

$$\frac{d}{dz}\begin{pmatrix} \mathscr{A}_1 \\ \mathscr{A}_{-1}^* \end{pmatrix} = \lambda \begin{pmatrix} \mathscr{A}_1 \\ \mathscr{A}_{-1}^* \end{pmatrix}, \tag{13}$$

the solution is

$$\lambda = \pm p_0 = \pm\sqrt{\beta(2\gamma - \beta)},$$
$$= \pm\sqrt{\gamma^2 - \frac{(\Delta k)^2}{4}}. \tag{14}$$

$\Delta k = 2(\gamma - \beta)$. When $\Delta k$ increases, this growth rate decreases. It was previously calculated by Bespalov and Talanov [19]. It is shown in [31] and Appendix A that $\Delta k$ is the phase mismatch between two cross coupled polarizations and the side-modes. Condition $\Delta k = 0$ is a resonance condition corresponding to a maximum energy transfer. An important consequence is that when the atmosphere is turbulent, the pump wave can be considered as the superposition of many waves, resulting in a spread of wave vectors [Appendix A]. The growth rate of the instability decreases as many wave vectors fail to meet the matching condition when the spread is large.

## 4) A simple calculation predicts the reduction in the growth rate of modulational instability due to turbulence

To further improve our understanding of how turbulence affects the development of modulational instability, we now describe a simple approach. The stochastic phase $\varphi$ is assumed to follow a process similar to the Kubo-Anderson process [20] which is set as a stepwise constant random function that jumps at randomly chosen distances between random step-values (Fig.2).

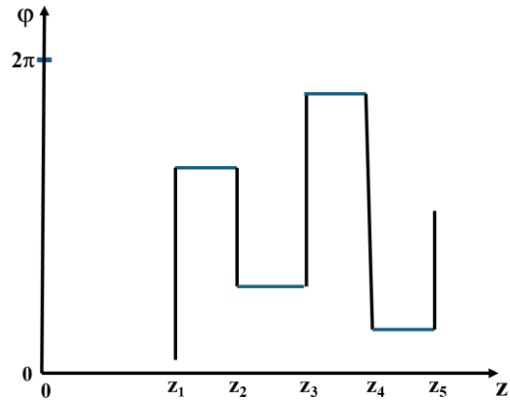

Fig.2. Evolution of the phase $j$ versus the propagation distance.

The values of $\varphi$ is a random number taken in the range [0-$2\pi$] and for the sake of simplicity it is supposed to jump every constant distance $1/\Delta\kappa$ [20] which is assumed to be small. Moreover, we consider that this distance is the spatial coherence length, $1/\Delta\kappa = 1/\Delta k$.

We introduce the two following quantities: $S = a_1 + a_{-1}$ and $D = a_1 - a_{-1}$, Eqs. (11) show that they must satisfy

$$\frac{d}{dz} S = i(\eta - \gamma)D, \qquad (15a)$$

$$\frac{d}{dz} D = i(\eta + \gamma)S, \qquad (15b)$$

with $\eta = \gamma - \beta + \chi\cos\varphi$. Thus

$$\frac{d^2}{dz^2} S = g^2 S, \qquad (16a)$$

$$\frac{d^2}{dz^2} D = g^2 D, \qquad (16b)$$

with

$$g^2 = -(\eta^2 - \gamma^2) = p_0^2 - \chi^2 \cos^2\varphi - 2(\gamma - \beta)\chi\cos\varphi, \tag{17}$$

where $g$ has the dimension of a growth rate.

Between two jumps $\varphi$ is constant and the solution for $S$ is given by solving Eq. (16a)

$$S = \lambda_1 \cosh(gz) + \lambda_2 \sinh(gz), \tag{18a}$$

and Eq. (15b) gives

$$D = i\left[\lambda_1 \sinh(gz) + \lambda_2 \cosh(gz)\right]\frac{1}{g}(2\gamma - \beta + \chi\cos\varphi). \tag{18b}$$

We have $\lambda_1 = S(0)$ and $\lambda_2 = -iD(0)g/(2\gamma - \beta + \chi\cos\varphi)$.

After a jump at $(j-1)/\Delta\kappa$, at the end of the following constant phase interval with $\varphi = \varphi_j$, expanding the two hyperbolic functions, the approximate values of S and D are given by

$$S\left(\frac{j}{\Delta\kappa}\right) = S\left(\frac{j-1}{\Delta\kappa}\right)\left(1 + \frac{1}{2}\frac{g^2}{\Delta\kappa^2}\right) - iD\left(\frac{j-1}{\Delta\kappa}\right)\frac{g}{\Delta k + \beta + \chi\cos\varphi_j}\frac{g}{\Delta\kappa},$$

$$D\left(\frac{i}{\Delta\kappa}\right) = iS\left(\frac{j-1}{\Delta\kappa}\right)\frac{g}{\Delta\kappa} + D\left(\frac{j-1}{\Delta\kappa}\right)\left(1 + \frac{1}{2}\frac{g^2}{\Delta\kappa^2}\right). \tag{19}$$

It is assumed that $\Delta\kappa \gg g$, which means that the distance between two jumps is much smaller than the distance required for the instability to significantly grow. In the case of low turbulence, it can be assumed that $\gamma \gg \chi$ and so $g$ is close to $p_0$ [Eq. (17)]. By performing iterations and calculating $S$ and $D$ for each phase step, considering that $\varphi$ is stochastic and $\Delta k = \Delta\kappa$, a dominant term can be identified for $S$.

$$S\left(\frac{n}{\Delta\kappa}\right) \approx S(0)\left(1 + \frac{1}{2}\frac{p_0^2}{\Delta\kappa^2}\right)^n, \tag{20}$$

where $n$ is an integer which goes to infinity. As $z = n/\Delta\kappa$

$$S(z) \approx S(0)\left(1 + \frac{1}{2}\frac{p_0^2}{\Delta\kappa^2}\right)^{z\Delta\kappa} = S(0)\left(1 + \frac{1}{2}\frac{p_0^2 z}{\Delta\kappa}\frac{1}{\Delta\kappa z}\right)^{z\Delta\kappa},$$

$$\approx S(0)\exp\left(\frac{1}{2}\frac{p_0^2}{\Delta\kappa}z\right), \tag{21}$$

while $D(z)$ has the same exponential behavior. The growth rate $p_0$ calculated when there is no turbulence is reduced by a factor $p_0/2\Delta\kappa$. A short distance between two phase jumps (large $\Delta\kappa$) will result in an even greater reduction of this growth rate.

## 5) A more sophisticated approach to the problem using the Fokker-Planck equation

Now, if we start again from Eqs (11). It is straight forward to verify that

$$\frac{\partial}{\partial z}\left[\left|\mathcal{A}_1\right|^2 - \left|\mathcal{A}_{-1}^*\right|^2\right] = 0, \tag{22}$$

then, the solution can be taken in the form

$$\begin{aligned}\mathcal{A}_1 &= \sqrt{q}\sinh\theta\, e^{i\varphi_1(z)}, \\ \mathcal{A}_{-1}^* &= \sqrt{q}\cosh\theta\, e^{i\varphi_2(z)},\end{aligned} \tag{23}$$

where $q$ is a constant and $\varphi_1$, $\varphi_2$ are two functions which depend on $z$.

When, $\varphi(z)$ and $\cos\varphi(z)$ are stochastic quantities. It is assumed that $\cos\varphi(z)$ undergoes a Brownian motion [23, 24]; it is a Langevin force with a correlation function proportional to a $\delta$ function [24]

$$\langle\cos\varphi(z)\cos\varphi(z')\rangle = \rho\delta(z-z'), \tag{24}$$

where $\rho$ is a constant.

Equations (11) and (23) lead to the Langevin equations for the system [23, 24]

$$\frac{d\theta}{dz} = -\gamma\sin\phi, \tag{25a}$$

$$\frac{d\phi}{dz} = -2(\gamma-\beta) - \gamma\cos\phi(\coth\theta + \tanh\theta) - 2\chi\cos\varphi, \tag{25b}$$

with $\phi = \varphi_2 - \varphi_1$.

In the absence of turbulence ($\chi = 0$), $\varphi_1$, $\varphi_2$ and therefore $\phi$ can be assumed to be three real constants. When $\theta \gg 1$ Eq; (25b) leads to $\cos\phi = -(\gamma-\beta)/\gamma$. Then, Eq. (25a) gives $\theta = \pm p_0 z$, which is again the growth rate of Bespalov and Talanov [19].

Now, a distribution function $f(\theta,\phi,z)$ is introduced. This function is assumed to be periodic in $\phi$. The equation of motion of this distribution is given by the Fokker-Planck equation

$$\frac{\partial f}{\partial z}+\left[\frac{\partial}{\partial \theta}D_\theta + \frac{\partial}{\partial \phi}D_\phi + D_{\theta\theta}\frac{\partial^2}{\partial \theta^2} + D_{\theta\phi}\frac{\partial^2}{\partial \theta \partial \phi} + D_{\phi\phi}\frac{\partial^2}{\partial^2\phi}\right] f = 0. \tag{26}$$

Coefficients $D_\theta$ and $D_\phi$ are the drift coefficients and $D_{\theta\theta}$, $D_{\theta\phi}$, $D_{\phi\phi}$ the diffusion coefficients.

The following general form can be used to express the Langevin equations.

$$\dot{x}_i = h_i(x,z) + g_{ij}(x,z)\Gamma_j(z), \tag{27}$$

where $x = (x_i)$ is a vector and the $\Gamma_j(z)$'s are Langevin forces. Assuming [24]

$$\langle \Gamma_i(z)\Gamma_j(z')\rangle = 2\delta_{ij}\delta(z-z'). \tag{28a}$$

and

$$\langle \Gamma_i(z)\rangle = 0 \tag{28b}$$

Considering these correlation functions we have [24]

$$D_i = h_i(x,z) + g_{kj}(x,z)\frac{\partial g_{ij}(x,z)}{\partial x_k},$$
$$D_{ij} = g_{ik}(x,z) g_{jk}(x,z). \tag{29}$$

The correlation functions given by (28a) are different from the one given by [24]. Still, constant $\rho$ in [24] can be absorbed into the initial functions $g_{ij}(x,z)$ by multiplying it by $\sqrt{\rho/2}$. In our case that is described by Eqs. (25), we have: $g_{\theta,\theta} = g_{\theta,\phi} = 0$ and $g_{\phi,\phi} = \sqrt{2\rho\chi}$. Then the drift and diffusion coefficients which determine the Fokker-Planck equation are given by

$$\begin{aligned}
D_\theta &= -\gamma \sin\phi, \\
D_\phi &= -2(\gamma-\beta) - \gamma\cos\phi(\coth\theta + \tanh\theta), \\
D_{\theta\phi} &= 0, \\
D_{\theta\theta} &= 0, \\
D_{\phi\phi} &= -2\rho\chi^2,
\end{aligned} \tag{30}$$

In the rest of this paragraph, we assume that: $\theta \gg 1$ (or $z \gg 1/p_0$). This implies $\coth\theta + \tanh\theta \approx 2$.

The Fokker-Planck equation for the distribution $f(\theta,\phi,z)$ reads

$$\frac{\partial f}{\partial z} - \gamma \sin\phi \frac{\partial f}{\partial \theta} + 2\gamma \sin\phi \, f - [2(\gamma - \beta) + 2\gamma \cos\phi]\frac{\partial f}{\partial \phi} = 2\rho\chi^2 \frac{\partial^2 f}{\partial \phi^2}. \tag{31}$$

The distribution $f$ is sought in the form of an expansion: $f = f_0 + f_1 + f_2...$, with $f_2 << f_1 << f_0$. It is assumed that the $f_i$'s are periodic in $\phi$ and to zero order: $\partial^2 f_0 / \partial \phi^2 = 0$. It implies that the zero-order distribution has the following form: $f_0 = f_0(\theta, z)$. To first order, when assuming that: $\gamma f_0 : \rho\chi^2 f_1$, $\partial f_0/\partial \theta \le f_0$, and $\partial f_0/\partial z$ is at most a first order quantity, the equation for the distribution function is

$$\frac{\partial f_0}{\partial z} - \gamma \sin\phi \frac{\partial f_0}{\partial \theta} + 2\gamma \sin\phi \, f_0 = 2\rho\chi^2 \frac{\partial^2 f_1}{\partial \phi^2}. \tag{32}$$

Integrating Eq (32) over $\phi$, as $f_0$ does not depend on $\phi$, $\partial f_0/\partial z$ has to be a second order term. We find

$$f_1 = \frac{\gamma \sin\phi}{2\rho\chi^2}\left(\frac{\partial f_0}{\partial \theta} - 2 f_0\right). \tag{33}$$

Equations (31) and (33) give the second order equation for $f$

$$\frac{\partial f_0}{\partial z} - \frac{\gamma^2 \sin^2\phi}{2\rho\chi^2}\left(\frac{\partial^2 f_0}{\partial \theta^2} - 2\frac{\partial f_0}{\partial \theta}\right) + \left(\frac{\partial f_0}{\partial \theta} - 2 f_0\right)\left\{\frac{\gamma^2 \sin^2\phi}{\rho\chi^2} - (\gamma - \beta + \gamma\cos\phi)\frac{\gamma\cos\phi}{\rho\chi^2}\right\} = 2\rho\chi^2 \frac{\partial^2 f_2}{\partial \phi^2}, \tag{34}$$

integrating this equation over $\phi$ between 0 and $2\pi$, we obtain

$$\frac{\partial f_0}{\partial z} + \frac{\gamma^2}{2\rho\chi^2}\frac{\partial f_0}{\partial \theta} = \frac{\gamma^2}{4\rho\chi^2}\frac{\partial^2 f_0}{\partial \theta^2}. \tag{35}$$

Introducing new variables: $\theta' = \theta - \frac{\gamma^2}{2\rho\chi^2} z$ and $\zeta = z$, equation (35) becomes

$$\frac{\partial f_0}{\partial \zeta} = \frac{\gamma^2}{4\rho\chi^2}\frac{\partial^2 f_0}{\partial \theta'^2}, \tag{36}$$

An elementary solution of this diffusion equation corresponding to an extreme concentration on the initial position is

$$f_0 = \left(\pi \frac{\gamma^2}{\rho\chi^2} z\right)^{-1/2} \exp\left[-\frac{\left(\theta - \frac{\gamma^2}{2\rho\chi^2} z - \theta'_0\right)^2}{\frac{\gamma^2}{\rho\chi^2} z}\right], \tag{37}$$

where $\theta'_0$ is a constant. The distribution (37) is a gaussian centered at $\theta = \theta'_0 + \left(\frac{\gamma^2}{2\rho\chi^2}\right)z$, with width $\sqrt{\left\langle\left(\theta - \theta'_0 - \frac{\gamma^2}{2\rho\chi^2}z\right)^2\right\rangle} = \frac{\gamma^2}{2\rho\chi^2}z$. Thus, the growth rate of the instability remains close to

$$p = \frac{\gamma^2}{2\rho\chi^2}, \tag{38}$$

and decreases at high values of $\chi$, that is to say, when the amplitude of $\delta n(z)$ is high.

We can find this result again by defining another distribution $F$ for $\phi$ only. The Fokker-Planck equation for the distribution $F(\phi)$ reads

$$\frac{\partial F}{\partial z} - 2\frac{\partial}{\partial \phi}\left[(\gamma - \beta + \gamma\cos\phi)F\right] = 2\rho\chi^2 \frac{\partial^2 F}{\partial \phi^2}. \tag{39}$$

Assuming that $\partial F/\partial z \approx 0$, we have

$$F = F_0 \exp-\left[\frac{(\gamma-\beta)\phi + \gamma\sin\phi}{\rho\chi^2}\right], \tag{40}$$

where $F_0$ is a constant. Consequently

$$\left\langle\frac{d\theta}{dz}\right\rangle = -\gamma\langle\sin\phi\rangle = -\gamma \frac{\int_0^{2\pi} \sin\phi \exp-\left[\frac{(\gamma-\beta)\phi + \gamma\sin\phi}{\rho\chi^2}\right]d\phi}{\int_0^{2\pi} \exp-\left[\frac{(\gamma-\beta)\phi + \gamma\sin\phi}{\rho\chi^2}\right]d\phi}, \tag{41}$$

by expanding the exponentials, we find.

$$\left\langle\frac{d\theta}{dz}\right\rangle = p \approx \frac{\gamma(2\beta - \gamma)}{2\rho\chi^2}. \tag{42}$$

This also implies that the growth rate decreases when turbulence is high.

We note that within the framework of this Fokker-Planck approach, when $\beta \approx \gamma$ which corresponds to a maximum growth rate in the non-turbulent case, we have

$$\left\langle \frac{d\theta}{dz} \right\rangle \approx p \approx p_0 \frac{p_0}{2\rho\chi^2}, \tag{43}$$

the growth rate of the Bespalov and Talanov instability is very much reduced when $\rho\chi^2 \gg p_0$.

Our conclusion is the same regardless of the approach taken: turbulence lowers the growth rate of the modulational instability, which could result in multiple filamentation.

## 6)  Conclusions

The influence of turbulence on the propagation of high-power laser pulses in the atmosphere, as well as its potential to induce multiple filamentation, has already been studied [9, 20, 23, 28, 32-34]. In this article, we have revisited the impact of turbulence on the multiple filamentation process when a laser beam propagates through the atmosphere [20]. Various analytical approaches were employed. The modulational instability of a plane wave is examined using the nonlinear Schrödinger equation, while a stochastic perturbation of the refractive index was introduced to model turbulence. Through multiple approaches and approximations, we have demonstrated that turbulence reduces the growth rate of the Bespalov-Talanov instability. At high power, turbulence is expected to suppress most multiple filamentation phenomena.

# Appendix A

*The influence of phase mismatch on the growth rate of modulational instability. A physical explanation for its reduction due to turbulence*

To gain insight into the physics of the problem, we present a physical approach that explains the mechanism behind the growth of the Bespalov-Talanov instability. Our focus is on the significance and necessity of satisfying a resonance condition. The objective is to demonstrate how turbulence can reduce the growth rate of this instability by making the resonance condition less strictly satisfied.

We consider the laser electric field to take the following form

$$\mathbf{E}(\mathbf{r},t) = \overline{\mathbf{E}}(\mathbf{r}) \exp(i\omega_0 t) + c.c. \quad (A1)$$

The wave equation reads

$$\left(\Delta - \frac{1}{c^2}\frac{\partial^2}{\partial t^2}\right)\mathbf{E}(\mathbf{r},t) = \mu_0 \frac{\partial^2}{\partial t^2}(\mathbf{P}_L + \mathbf{P}_{NL}), \quad (A2)$$

where $\mathbf{P}_{NL}$ is the third-order nonlinear polarization which can play a role in both centrosymmetric and noncentrosymmetric media [25].

We have

$$\mathbf{P}_{NL} = \varepsilon_0 \chi^{(3)} E^3(r,t), \quad (A3)$$

where $\chi^{(3)}$ is the nonlinear susceptibility. Keeping the part at frequency $\omega_0$ only, we find

$$\mathbf{P}_{NL} = 3\varepsilon_0 \chi^{(3)}\left[\left|\overline{\mathbf{E}}\right|^2 \overline{\mathbf{E}} \exp(i\omega_0 t)\right]. \quad (A4)$$

The electric field is assumed to be the sum of three terms

$$\mathbf{E}(\mathbf{r},t) = \mathbf{E}_0(\mathbf{r},t) + \mathbf{E}_1(\mathbf{r},t) + \mathbf{E}_{-1}(\mathbf{r},t), \quad (A5)$$

where $\mathbf{E}_1(\mathbf{r},t)$ and $\mathbf{E}_{-1}(\mathbf{r},t)$ represent two weak symmetrical side modes.

The part of the nonlinear polarization which is phase-matched to the strong component is

$$\mathbf{P}_{0NL} = 3\varepsilon_0 \chi^{(3)}\left|\overline{\mathbf{E}}_0\right|^2 \overline{\mathbf{E}}_0, \quad (A6)$$

and the part that is phase matched to the sidemodes in the transverse direction is given by

$$\mathbf{P}_{\pm NL} = 3\varepsilon_0 \chi^{(3)}\left[2\left|\overline{\mathbf{E}}_0\right|^2 \overline{\mathbf{E}}_{\pm 1} + \overline{\mathbf{E}}_0^2 \overline{\mathbf{E}}_{m1}^*\right]. \quad (A7)$$

As the propagation equation [Eq. (10)] is obtained by setting to zero the terms in $e^{i\mathbf{k}_\perp \cdot \mathbf{r}}$ and in $e^{-i\mathbf{k}_\perp \cdot \mathbf{r}}$, $\mathbf{E}_1(\mathbf{r},t)$ and $\mathbf{E}_{-1}(\mathbf{r},t)$ must be phase matched to $\overline{\mathbf{E}}_0^2 \overline{\mathbf{E}}_{-1}^*$ and $\overline{\mathbf{E}}_0^2 \overline{\mathbf{E}}_{+1}^*$ respectively.

When there is no turbulence, the wave equation for the off-axis mode can be put in the following form

$$\frac{\partial a_{\pm 1}}{\partial z} = -i\beta a_{\pm 1} + 2i\gamma a_{\pm 1} + i\gamma a_{m1}^{*} e^{2i\gamma z}. \tag{A8}$$

On the right-hand side of this equation the second term represents the cross-phase modulation from the pump wave to the signal waves [25, 31, 35] This term contains the additional lengthening of the weak waves propagation vectors in the z-direction; the nonlinear index for the weak waves is twice the one of the high intensity wave [25]. The last term describes the energy transfer from the pump to the signal waves. The total wave vector of the sidemodes is: $k_{oT} = k_0 + 2\gamma - \beta$. Thereby, we search for a solution for Eqs. (A10) in the form.

$$a_{\pm 1} = \hat{a}_{\pm 1} exp\left[i(2\gamma - \beta)z\right], \tag{A9}$$

where $\hat{a}_1$ and $\hat{a}_{-1}$ must satisfy the following equations [31]

$$\frac{d^2}{dz^2}\hat{a}_{\pm 1} + i\Delta k \frac{d}{dz}\hat{a}_{\pm 1} - \gamma^2 \hat{a}_{\pm 1} = 0. \tag{A10}$$

Solving Eqs. (A12) we obtain the solution for $\hat{a}_{\pm 1}$.

$$\hat{a}_{\pm 1} : exp\left[\sqrt{\left(\gamma^2 - \frac{(\Delta k)^2}{4}\right)}z\right] exp\left(-i\frac{\Delta k}{2}z\right). \tag{A11}$$

Using Eqs. (A9), we have

$$a_{\pm 1} : exp\left[\sqrt{\left(\gamma^2 - \frac{(\Delta k)^2}{4}\right)}z\right] exp\left(-i\frac{\Delta k}{2}z\right) exp\left[i(2\gamma - \beta)z\right]. \tag{A12}$$

This confirms that the growth rate of the two signal waves decreases when $\Delta k$ increases. The maximum is reached when. $\Delta k = 0$.

Now, let us give more physical meaning to $\Delta k$. To do so, it is possible to consider Eq. (A12) over a long space scale. The right-hand side contains two first exponentials which remain almost constant when the conditions $\Delta k \approx 0$ and $\gamma = 1$ are satisfied. Therefore, only the third exponential undergoes rapid oscillations. Thus, it can be considered that the pump wave and its two radial signals, when taking into consideration the nonlinear refractive index of the pump wave, are represented in the following form

$$\begin{aligned}
\mathbf{E}_0 &= \tilde{E}_0 exp\left(ik_{0T}z - i\omega_0 t\right), \\
\mathbf{E}_{\pm 1} &= \tilde{E}_{\pm 1} exp\left(ik_{\pm 1}z - i\omega_0 t\right),
\end{aligned} \tag{A13}$$

with $k_{0T} = k_0 + \gamma$, $k_{\pm 1} = k_0 + 2\gamma - \beta$ ($k_1 = k_{-1}$) and where the $\tilde{E}_0$ and $\tilde{E}_{\pm 1}$ are slowly varying quantities. The instability grows when some resonance condition is met. The mismatch of $\mathbf{E}_{+1}$ and $\mathbf{E}_{-1}$ with their corresponding cross-coupled polarizations $\mathbf{E}_0^2 \mathbf{E}_{-1}^*$ and $\mathbf{E}_0^2 \mathbf{E}_{+1}^*$ [Eq. (A7)] is [25, 31, 35]

$$\Delta \tilde{k} = 2k_{0T} - 2k_{\pm 1} = -\Delta k. \tag{A14}$$

The coupling is maximized when the mismatch $\Delta \tilde{k}$ or $\Delta k$ is zero. In this case, we once again find that the growth rate is maximum when $\Delta k = 0$.

In a turbulent atmosphere, the wave can be viewed as the superposition of multiple waves with different wave vectors corresponding to different refractive indexes. The Fourier expansion of the wave represents the sum of these multiple realizations, leading to a spread of wave vectors. When this spread is large, many of these wave vectors fail to satisfy the matching condition, resulting in a decreased growth rate.